\documentclass{AN_ART}
\usepackage{graphicx}
\usepackage{amsmath}
\begin{document}
\begin{titlepage}
\setcounter{page}{001} \headnote{Astron.~Nachr.~322 (2001) 3,
001--010}
\makeheadline
\title{On the stability of compact supermassive objects}
\author{{\sc L.~Verozub},
Kharkov, Ukraine \\
\medskip
{\small Kharkov National University} \\
\bigskip
{\sc A.~Kochetov}, Kharkov, Ukraine\\
\medskip
{\small Kharkov National University} \\
}
\date{Received  2000 April 19; accepted 2001 June 29}
\maketitle
\summary Proceeded from the gravitation equations proposed by one
of authors it was argued in a previous paper that there can exist
supermassive compact configurations of degenerated Fermi-gas
without events horizon. In the present paper we consider the
stability of these objects by method like the one used in the
theory of stellar structure. It is shown that the configurations
with an adiabatic equation of state with the power $\geq 4/3$ are stable.\\
END
\keyw compact objects --- massive objects
END
\end{titlepage}
\section{Introduction}

Thirring [\cite{Thirring}] proposed that gravitation could be described as a
tensor field $\psi_{\alpha\beta}(x)$ of spin two in 4 - dimensional
Pseudo-Euclidean space-time $E_{4}$, where the Lagrangian, describing
the motion of test particles in a given field, is of the form
\begin{equation}
L=-m_{p}c\left[  g_{\alpha\beta}(\psi)\dot{x}^{\alpha}\dot{x}^{\beta}\right]
^{1/2}\;. \label{Thirring_Lagr}%
\end{equation}
In this equation $g_{\alpha\beta}$ is a tensor function of $\psi_{\alpha\beta
}$, $m_{p}$ is the mass of the particle, $c$ is the speed of light and
$\dot{x}^{\alpha}=dx^{\alpha}/dt$ (Greek indices run from 0 to 3).

A theory based on such a Lagrangian must be invariant under some gauge
transformations $\psi_{\alpha\beta}\longrightarrow\bar{\psi}_{\alpha\beta}$
that are a consequence of the existence of  ''extra'' components of the tensor
$\psi_{\alpha\beta}$. Transformations $\psi_{\alpha\beta}\longrightarrow
\bar{\psi}_{\alpha\beta}$ give rise to transformations $g_{\alpha\beta}$
$\longrightarrow$ $\bar{g}_{\alpha\beta}$. Therefore, the field equations for
$g_{\alpha\beta}(x)$ and equations of the motion of a test particle must be
invariant under these transformations of the tensor $g_{\alpha\beta}$.

The equations of motion of the test particle, resulting from eq.
(\ref{Thirring_Lagr}) are also equations of geodesic lines of the
4-dimensional Riemannian space-time $V_{n}$ whose the metric tensor is
$g_{\alpha\beta}(\psi)$. Therefore, if the transformation $\psi_{\alpha\beta
}\longrightarrow\bar{\psi}_{\alpha\beta}$ leaves the equation of the motion
invariant, then the corresponding transformation $g_{\alpha\beta}$
$\longrightarrow$ $\bar{g}_{\alpha\beta}$ is some mapping $V\longrightarrow
\bar{V}$ of the Riemannian spaces leaving geodesic lines invariant, i.e. a
geodesic (projective) one. If not only eq. (\ref{Thirring_Lagr}) but also the
equations of field contain $\psi_{\alpha\beta}$ in the form $g_{\alpha\beta
}(\psi)$, then clear that only geodesic-invariant equations for $g_{\alpha
\beta}$ are permissible in such a theory. Such of kind
equations which have not physical singularity in the spherically-symmetric
field was proposed in the paper [\cite{Verozub1}]. These equations are of the form%

\begin{equation}
B_{\alpha\beta;\gamma}^{\gamma}-B_{\alpha\delta}^{\epsilon}B_{\beta\epsilon
}^{\delta}=0. 
\label{myeqs}%
\end{equation}
The equations are vacuum bimetric equations for the tensor
\begin{equation}
B_{\alpha\beta}^{\gamma}=\Pi_{\alpha\beta}^{\gamma}-\overset{\circ}{\Pi
}_{\alpha\beta}^{\gamma}, \label{tensB}%
\end{equation}
where
\begin{equation}
\Pi_{\alpha\beta}^{\gamma}=\Gamma_{\alpha\beta}^{\gamma}-(n+1)^{-1} \left[
\delta_{\alpha}^{\gamma}\Gamma_{\epsilon\beta}^{\epsilon}+\delta_{\beta
}^{\gamma}\Gamma_{\epsilon\alpha}^{\epsilon}\right]  , \label{Thomases}%
\end{equation}%
\begin{equation}
\overset{\circ}{\Pi}_{\alpha\beta}^{\gamma}=\overset{\circ}{\Gamma}%
_{\alpha\beta}^{\gamma}-(n+1)^{-1}\left[  \delta_{\alpha}^{\gamma}%
\overset{\circ}{\Gamma}_{\epsilon\beta}^{\epsilon}+\delta_{\beta}^{\gamma
}\overset{\circ}{\Gamma}_{\epsilon\alpha}^{\epsilon}\right]  ,
\label{Thomases0}%
\end{equation}
$\overset{\circ}{\Gamma}_{\alpha\beta}^{\gamma}$ are the Christoffel symbols
of space-time $E_{4}$ whose fundamental tensor is $\eta_{\alpha\beta}$,
$\Gamma_{\alpha\beta}^{\gamma}$ are the Christoffel symbols of the Riemannian
space-time $V_{4}$, whose fundamental tensor is $g_{\alpha\beta}$ and
$\delta_{\beta}^{\gamma}$ is the Kroneker delta. The
semi-colon in eq. (\ref{myeqs}) denotes the covariant differentiation in
$E_{4}$ \footnote{There was a misprint in eqs. (9) of the paper
[\cite{Verozub2}]. These equations must be read as follows:\newline
\begin{equation}
\lim\limits_{r\rightarrow\infty}A=1, \; \lim\limits_{r\rightarrow\infty}%
(B/r^{2})=1, \;\lim\limits_{r\rightarrow\infty}C=1. \label{limitConditions}%
\end{equation}
}.

Eqs. (\ref{myeqs}) are invariant under arbitrary transformations of the tensor
$g_{\alpha\beta}$ retaining invariant the equations of motion of a test
particle. Thus, the tensor field $g_{\alpha\beta}$ is defined up to geodesic
mappings of space-time $V_{4}$ (in the analogous way as the potential
$A_{\alpha}$ in electrodynamics is determined up to gauge transformations). A
physical sense in the theory have only geodesic invariant values. The simplest
object of that kind is the object $B_{\alpha\beta}^{\gamma}$ which can be
named the strength tensor of gravitation field. The coordinate system is
defined by the used measurement instruments and in each case is a given.

Because of the gauge (i.e. geodesic or projective) invariance, additional
conditions can be imposed on the tensor $g_{\alpha\beta}$. In particular
[\cite{Verozub1}], under the condition
\begin{equation}
Q_{\alpha}=\Gamma_{\alpha\sigma}^{\sigma}-\overset{\circ}{\Gamma}%
_{\alpha\sigma}^{\sigma}=0 \label{gaugeConditions}%
\end{equation}
eqs. (\ref{myeqs}) are equivalent to the following system:%

\begin{equation}
R_{\alpha\beta}=0, \label{Einstein Eqs}%
\end{equation}%
\begin{equation}
Q_{\alpha}=0, \label{AditionalCondition}%
\end{equation}
where $R_{\alpha\beta}$ is the Ricci tensor.

Proceeded from these equations in the papers [\cite{Verozub2}] --
[\cite{Verozub4}], it was shown that there can exist supermassive equilibrium
compact objects without events horizon. Maybe just such of kind objects are in
the galactic centers [\cite{Verozub5}].

A geodesic-invariant generalization of the equations (\ref{myeqs}) inside
matter can be found in the following way (see also [\cite{VerKoch2000}]).

Transformations of the Christoffel symbols under the geodetic (i.e.
projective) mappings is of the form:%

\begin{equation}
\overset{\_}{\Gamma}_{\beta\gamma}^{\alpha}=\Gamma_{\beta\gamma}^{\alpha
}+\varphi_{\beta}\delta_{\gamma}^{\alpha}+\varphi_{\gamma}\delta_{\beta
}^{\alpha}, \label{ChristoffelTransformation}%
\end{equation}
where $\varphi_{\gamma}$ is a vector-function of $x^{\alpha}$. This equation
has a simple interpretation in an 5- dimensional manifold $\mathcal{M}_{5}$,
where the admissible coordinates transformations are given by%

\begin{equation}
\overset{\_}{x}^{\alpha}=\overset{\_}{x}^{\alpha}(x^{0},x^{1},x^{2},x^{3}),
\label{5coordTransform1}%
\end{equation}%

\begin{equation}
\overset{\_}{x}^{4}=x^{4}-\int\varphi_{\alpha}dx^{\alpha}.
\label{5coordTransform2}%
\end{equation}
Namely, eqs. (\ref{ChristoffelTransformation})\ can be interpreted as the
transformation of 4-components $\Gamma_{\beta\gamma}^{\alpha}$ of the
connection coefficient $\Gamma_{BC}^{A}$ ($A,B,C=0..4$) in $\mathcal{M}_{5}$
under the transformation (\ref{5coordTransform2}), if the components
$\Gamma_{4\beta}^{\alpha}$ obey the condition $\Gamma_{4\beta}^{\alpha}%
=\delta_{\beta}^{\alpha}$ .

For this reason we will consider the tensor $g_{\alpha\beta}$ as 4-components
of $n+1$- dimensional tensor
\begin{equation}
g_{AB}=\left(
\begin{array}
[c]{cc}%
g_{\alpha\beta} & g_{\alpha4}\\
g_{4\alpha} & g_{44}%
\end{array}
\right)  . \label{5dimensionMetricTensor}%
\end{equation}

The components $g_{\alpha\beta}$ are transformed under (\ref{5coordTransform2}%
) as follows:%

\begin{equation}
\overset{\_}{g}_{\alpha\beta}=g_{\alpha\beta}+g_{44}\varphi_{\alpha}%
\varphi_{\beta}+g_{4\alpha}\varphi_{\beta}+g_{4\beta}\varphi_{\alpha},
\label{galphabetaTransform}%
\end{equation}%

\begin{equation}
\overset{\_}{g}_{\alpha4}=g_{\alpha4}+g_{44}\varphi_{\alpha},
\label{g4alphaTransform}%
\end{equation}%

\begin{equation}
\overset{\_}{g}_{44}=g_{44}. \label{g44Transform}%
\end{equation}

Transformations of the components $\Gamma_{\alpha\beta}^{\beta}$ under the
geodesic mappings are given by%

\[
\overline{\Gamma}_{\alpha\beta}^{\beta}=\Gamma_{\alpha\beta}%
^{\beta}+(n+1)\psi_{\alpha},
\]
where $n$ is dimension of our space-time. Therefore, such a
transformation for $Q_{\alpha}$ coincides with
(\ref{g4alphaTransform}), if $g_{44}=n+1.$ For this reason, we
will assume that
\begin{equation}
g_{AB}=\left(
\begin{array}
[c]{cc}%
g_{\alpha\beta} & Q_{\alpha}\\
Q_{\alpha} & n+1
\end{array}
\right)  .\label{g_AB}%
\end{equation}

Then there exists the geodesic-invariant tensor
\begin{equation}
G_{\alpha\beta}=g_{\alpha\beta}-(n+1)^{-1}Q_{\alpha}Q_{\beta},
\label{5dimMetricTensorInvatiant}%
\end{equation}
and the geodesic-invariant generalization of the Einstein equations with
matter source are of the form%

\begin{equation}
B_{\alpha\beta\>;\>\gamma}^{\gamma}-B_{\alpha\sigma}^{\gamma}B_{\beta\gamma
}^{\sigma}=k\left(  T_{\alpha\beta}-1/2G_{\alpha\beta}T\right)
,\label{MainEquationWithSource}%
\end{equation}
where $k=8\pi G/c^{4}$, $G$ is the gravitational constant, $c$ is speed of
light, $T_{\alpha\beta}$ is the matter energy-momentum tensor, $T=G^{\alpha
\beta}T_{\alpha\beta}$. Thus, we assume that inside matter the gravitation
equations under consideration just as the vacuum equations coincide with the
Einstein equations at the gauge conditions $Q_{\alpha}=0$.

In the previous paper [\cite{Verozub2}] the objects was considered as
homogeneous. In the present paper we find the solution of the gravitation
equations inside the objects and obtain more rigorous prove of the objects
stability.
%

\section{The Internal Structure}

Consider here the relativistic equations of the objects structure. In the
spherically symmetric field the metric differential form of space-time $V_{4}$
is given by [\cite{Verozub1}]
\begin{equation}
ds^{2}=C\,{dx^{0}}^{2} -A\,dr^{2}-
B\,[d\theta^{2}+\sin^{2}\theta\, d\varphi^{2}]%
\label{ds^{2}}
\end{equation}
where $A$, $B$ and $C$ are the functions of the radial coordinate $r$.

Temporarily let us replace in eq. (\ref{ds^{2}}) the radial coordinate $r$ by
$f=\sqrt{B}$:
\begin{equation}
ds^{2}=e^{\lambda}\,{dx^{0}}^{2}-e^{\gamma}\,df^{2}-f^{2}(d\theta^{2}+\sin
^{2}\theta \,d\varphi^{2}),
\end{equation}
where
\begin{equation}
e^{\lambda}=C,\,\,e^{\gamma}=\frac{A}{d f/d r}.
\end{equation}

We must solve the system of the equations
\begin{equation}
R_{\alpha\beta}=k(T_{\alpha\beta}-\frac{1}{2}g_{\alpha\beta}%
T),\label{einst_eqs}%
\end{equation}%
\begin{equation}
Q_{\alpha}=0.
\end{equation}

For an ideal fluid eqs. (\ref{einst_eqs}) are
\[
e^{-\gamma}\left(  \frac{\lambda_{f}^{^{\prime}}}{f}+\frac{1}{f^{2}}\right)
-\frac{1}{f^{2}}=kp,
\]%
\begin{equation}
e^{-\gamma}\left(  \frac{\gamma_{f}^{^{\prime}}}{f}-\frac{1}{f^{2}}\right)
+\frac{1}{f^{2}}=k\mu,
\end{equation}%
\[
\frac{e^{-\gamma}}{2}\left(  \lambda_{ff}^{\prime\prime}+\frac{\lambda
_{f}^{\prime2}}{2}-\frac{\lambda_{f}^{^{\prime}}\gamma_{f}^{\prime}}{2}%
+\frac{\lambda_{f}^{\prime}}{f}-\frac{\gamma_{f}^{\prime}}{f}\right)  =kp,
\]
where $\lambda_{f}^{\prime}=d\lambda/df$,
$\gamma_{f}^{\prime}=d\gamma/df$,
$\lambda_{ff}^{\prime\prime}=d^{2}\lambda/df^{2},$ $\mu=\rho
c^{2}$, $\rho$ is the matter density and $p$ is the pressure.

These equations are equivalent to the following system [\cite{Tolman}]:
\[
e^{-\gamma}\left(  \frac{\lambda_{f}^{^{\prime}}}{f}+\frac{1}{f^{2}}\right)
-\frac{1}{f^{2}}=kp,
\]%
\begin{equation}
e^{-\gamma}\left(  \frac{\gamma_{f}^{^{\prime}}}{f}-\frac{1}{f^{2}}\right)
+\frac{1}{f^{2}}=k\mu,\label{GeneralSolution2}%
\end{equation}%
\[
\frac{d p}{d f}=-\frac{1}{2}\left(  \mu+p\right)
\lambda_{f}^{\prime}.
\]
The general solution of this system is given by
\begin{equation}
e^{\lambda}=\sigma_{1}\left(  1-\frac{\Psi+\sigma_{2}}{f}\right)  e^{\Phi
},\label{S1}%
\end{equation}%
\begin{equation}
e^{\gamma}={\left(  1-\frac{\Psi+\sigma_{2}}{f}\right)}^{-1},
\label{S10}%
\end{equation}
\begin{equation}
\frac{dp}{df}=-\frac{1}{2}\frac{(\mu+p)}{f^{2}}\frac{\left(  \Psi+\sigma
_{2}+kpf^{3}\right)  }{1-(\Psi+\sigma_{2})/f},\label{S1a}%
\end{equation}
where
\begin{equation}
\Psi=k\int\mu f^{2}df,
\end{equation}%
\begin{equation}
\Phi=k\int\frac{\left(  \mu+p\right)  fdf}{1-(\Psi+\sigma_{2})/f},
\end{equation}
$\sigma_{1}$ and $\sigma_{2}$ are constants. Returning to the variable $r$ we
obtain finally three equations:
\[
C=\sigma_{1}\left(  1-\frac{\Psi+\sigma_{2}}{f}\right) e^{\Phi},
\]%
\begin{equation}
\left(  \frac{df}{dr}\right)  ^{2}=A\left(  1-\frac{\Psi+\sigma_{2}}%
{f}\right)  ,\label{eqs_star_general}%
\end{equation}%
\[
\frac{dp}{dr}=-\frac{1}{2}\frac{(\mu+p)}{f^{2}}\frac{\left(  \Psi+\sigma
_{2}+kpf^{3}\right)  }{1-(\Psi+\sigma_{2})/f}\frac{df}{dr}.
\]
The gauge condition (\ref{gaugeConditions}) yields 4th equation:
\begin{equation}
A C f^{4}r^{-4}=\sigma_{3},\label{add_cond_eqs}%
\end{equation}
where $\sigma_{3}$ is a constant.

The eqs. (\ref{eqs_star_general}) and (\ref{add_cond_eqs}) are the general
solution of the used gravitation equations inside and outside objects under consideration.

The constant $\sigma_{3}$ because of the conditions (\ref{limitConditions}) is
equal to 1. The constants $\sigma_{1}$ and $\sigma_{2}$ for the solution
outside the considered objects can be found from the  conditions at the
infinity (\ref{limitConditions}). 
Setting in eq. (\ref{eqs_star_general})
$\mu=0$ and  $p=0$ we obtain equations%

\begin{equation}
C =\sigma_{1}\left(  1-\frac{\sigma_{2}}{f}\right)
,\label{exp_Lambda}%
\end{equation}%
\begin{equation}
\left(\frac{df}{dr}\right)^{2} = A \left(  1-\frac{\sigma_{2}}{f}\right)  ,\label{F_prime}%
\end{equation}
which together with (\ref{add_cond_eqs}) yields the
solution of the vacuum gravitation equations (\ref{myeqs}) [\cite{Verozub1}]. 
The
constants are: $\sigma _{1}=1$, $\sigma_{2}=r_{g}=2 G M / c^{2}$,
where $M$ is the mass of the considered object.

Consider the solution of eqs. (\ref{eqs_star_general}) inside the
objects. From the condition that at $r=R$, where $R$ is the radius
of the object, the right hands of eqs. (\ref{exp_Lambda}) and
(\ref{F_prime}) must coincide with the ones of the first two eqs.
(\ref{eqs_star_general}) we find%

\begin{equation}
\sigma_{1} =\exp[-\Psi_{|f=f(R)}],
\end{equation}%
\begin{equation}
\sigma_{2} =r_{g}-\Phi_{|f=f(R)}.
\end{equation}

By using the notions
\begin{equation}
u=\Psi_{|f=f(R)}-\Psi,\label{c}%
\end{equation}%
\begin{equation}
v=\exp\left[  \frac{\Phi_{|f=f(R)}-\Phi}{2}\right]  ,\label{d}%
\end{equation}
we obtain finally the following system of the equation which determine the
internal structure of the object:
\begin{equation}
\frac{df}{dr}=\frac{r^{2}}{f^{2}}v,\label{eqs_inside_star_final}%
\end{equation}%
\begin{equation}
\frac{dp}{dr}=-\frac{1}{2}\frac{(\mu+p)r^{2}}{f^{3}}\,\frac{kpf^{3}+r_{g}%
-u}{f-r_{g}+u}\,v,\label{e}%
\end{equation}%
\begin{equation}
\frac{du}{dr}=-k\mu r^{2}v,\label{f}%
\end{equation}%
\begin{equation}
\frac{dv}{dr}=-\frac{k}{2}\frac{(\mu+p)r^{2}}{f-r_{g}+u}\,v^{2}\label{g}%
\end{equation}
and the boundary conditions:
\begin{equation}
f(R)=(R^{3}+r_{g}^{3})^{1/3},\quad p(R)=0,\quad u(R)=0,\quad v(R)=1.
\end{equation}

The function $v$ very little differs from 1 because the power of the exponent
(\ref{d}) is much less than 1. Therefore, it follows from the
(\ref{eqs_inside_star_final}) that the  function $f$ very little differ from
the ones in the solution of the vacuum equations (\ref{myeqs}).

At $R\gg r_{g}$ the function $f \simeq r$. At this condition these equations
coincide with the ones of general relativity.

We use an approximation equation by Harrison [\cite{Harrison}] for the baryons
density $n_{b}$ as a function of the matter density $\rho$ which is right
from $8$ to at least $10^{14}$ $g/cm^{3}$ :
\begin{equation}
n_{b}=q_{1}\rho(1+q_{2}\rho^{9/16})^{-4/9},\label{Harrison_state_equations}%
\end{equation}
where $q_{1}=6.0228\cdot10^{23}$ and $q_{2}=7.7483\cdot10^{-10}$ in CGS units,
and the equation of the state
\begin{equation}
p=\left(  \frac{n_{b}}{\partial n_{b}/\partial\rho}-\rho\right)
c^{2}.\label{EquationOfState}%
\end{equation}

\begin{figure}
\resizebox{115mm}{!}
{\includegraphics[width=115mm,height=60mm]{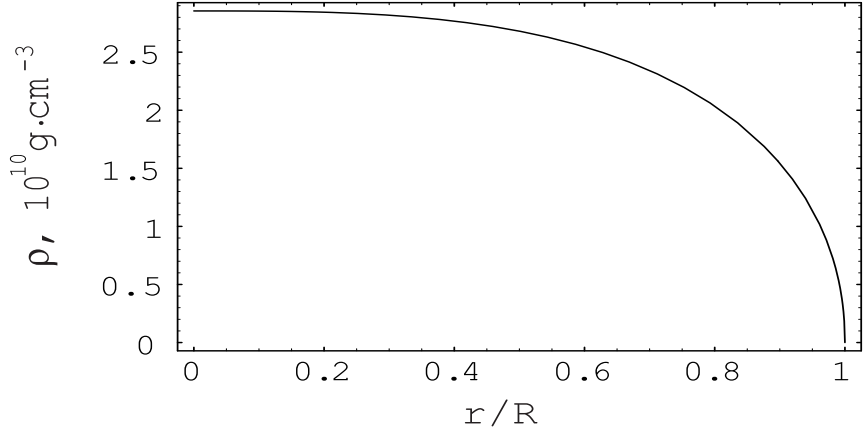}}
\hfill
\parbox[b]{55mm}{\caption{The function $\rho(r)$ for the configuration
with the mass $2.6\cdot10^{6}M_{\odot}$}
\label{density_of_r}}
\end{figure}

In addition to the ordinary solution of eq.
(\ref{eqs_inside_star_final})(i. e. configurations of the white
drafts and neutron stars) there exist solutions with large masses
from $10^{2}M_{\odot}$ up to $10^{10}M_{\odot}$. Fig.
\ref{density_of_r}\thinspace\ show the distribution of the density
$\rho(r)$ inside the configuration with the mass
$2.6\cdot10^{6}M_{\odot}$ (its radius is equal to
$0.057R_{\odot})$.

\begin{figure}
\resizebox{115mm}{!}
{\includegraphics[width=115mm,height=60mm]{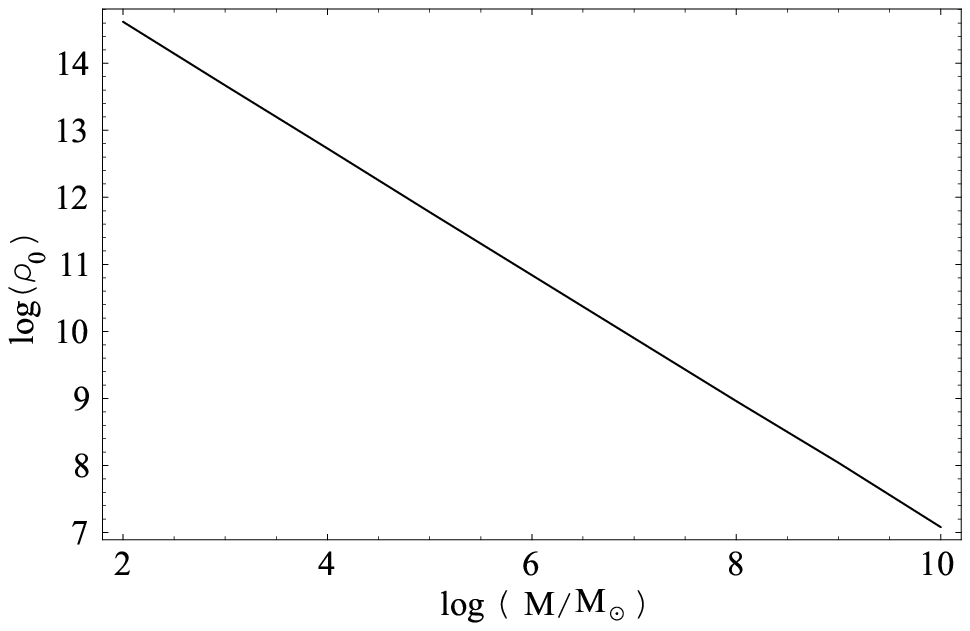}}
\hfill
\parbox[b]{55mm}{\caption{The relation between the central dencity and the
object radius}
\label{Rho_M}}
\end{figure}

\begin{figure}
\resizebox{115mm}{!}
{\includegraphics[width=115mm,height=60mm]{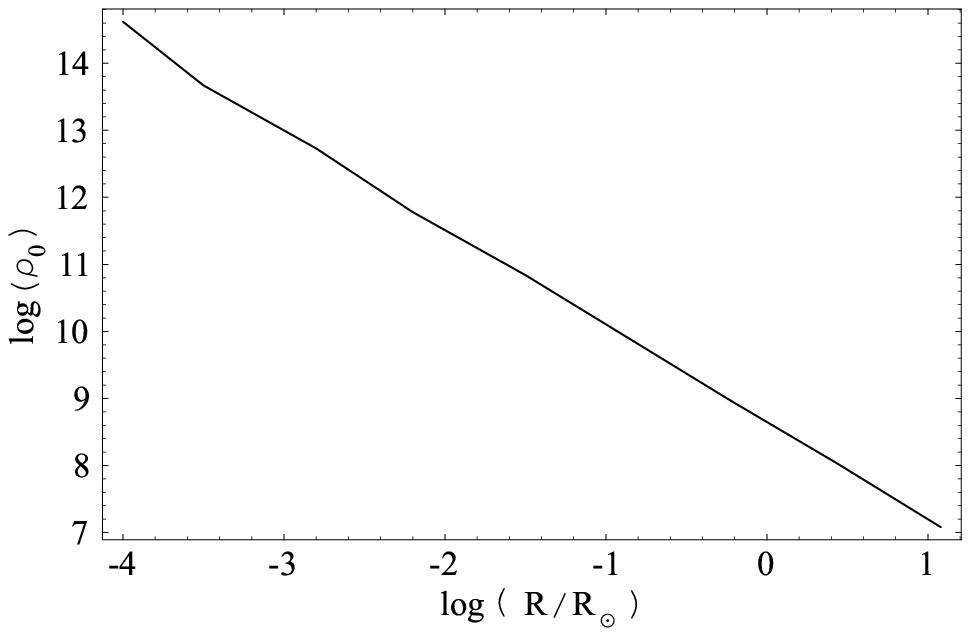}}
\hfill
\parbox[b]{55mm}{\caption{The relation between the central density and the
object mass}
\label{Rho_R}}
\end{figure}

\begin{figure}
\resizebox{115mm}{!}
{\includegraphics[width=115mm,height=60mm]{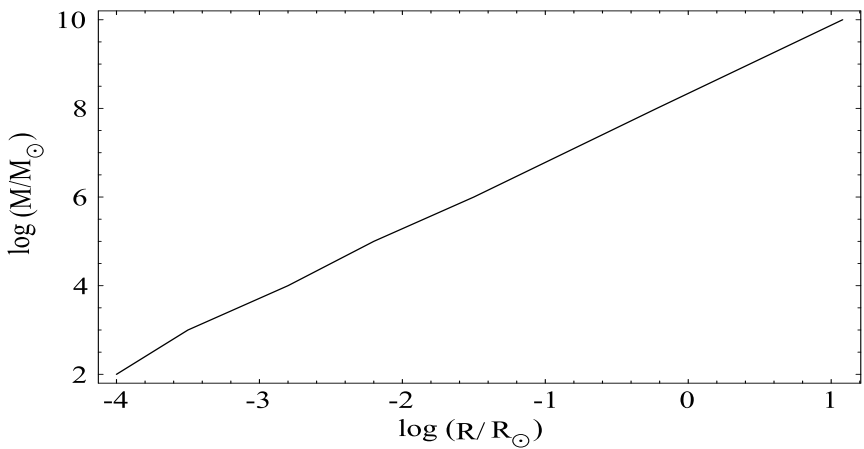}}
\hfill
\parbox[b]{55mm}{\caption{The relation between the masses and radiuses of the
objects}
\label{R_M}}
\end{figure}

Figs. \ref{Rho_M}, \ref{Rho_R} and \ref{R_M} show the relations  
" central density-- mass" , "central density--radius" and "mass-radius"
for the objects under consideration.

There are no solutions for greater masses. 

In according to figure \ref{R_M} the relation ''mass-radius'' for these 
configuration is given
by :
\begin{equation}
\frac{R}{R_{\odot}}=4.07\cdot10^{-6}\left(  \frac{M}{M_{\odot}}\right)
^{0.647} .%
\end{equation}
%

\section{Stability of Equilibrium Configurations}

It is not clear, at present, whether the found solutions for the big
masses make up a continuous class of solutions with the solutions for
neutron stars and dwarfs or not. 
Besides, the above figures have not any extremes points. For this reason
the figures do not give us direct evidences of the stability or instability
of the objects under consideration.

The stability of the objects was argued first
in paper [\cite{Verozub2}] where the objects was considered as
homogeneous. 
Following a classical method by [\cite{Ledoux}] and [\cite{Cox}] 
consider now
the problem of the stability of the configurations more rigorously, without
the assumption of homogeneity.

The complete energy of a spherically symmetric object of the mass $M$ and the
radius $R$ can be written as follows%

\begin{equation}
E=\int\limits_{0}^{M}\left(  u+\phi\right)  dm,\label{Pouwer}%
\end{equation}
where $m$ is the mass of the matter inside the sphere of the radius $r,$ $u$
is the intrinsic energy, $\phi$ is the gravitational potential. The
function $\mathcal{L}=u+\phi$ will be considered as the function of the
variables $m$, $r(m)$ and $r^{\prime}(m)=dr/dm$. Consider the functional%

\begin{equation}
E=\int\limits_{0}^{M}\mathcal{L}\left[ m,r(m),r^{\prime}(m)\right]
dm
\label{Power1}%
\end{equation}
on the set of all continuously differentiable function $r(m)$ with the
boundary conditions $r(0)=0$, $r(M)=R$ and find at which conditions the first
variation $\delta E$ is equal to zero and the second one is a positive at
small isoentropic disturbances $r\longrightarrow r+\delta r$.

Since
\begin{equation}
r^{\prime}=(4\pi\rho r^{2})^{-1},\quad\frac{\partial}{\partial r}=-\frac
{2\rho}{r}\frac{\partial}{\partial\rho},\quad\frac{\partial}{\partial
r^{\prime}}=-4\pi\rho^{2}r^{2}\frac{\partial}{\partial\rho}, \label{Variables}%
\end{equation}
the condition $\delta E = 0$ yields the Lagrange equation and end condition.

The Lagrange equation
\begin{equation}
\frac{d}{dm}\left(  \frac{\partial\mathcal{L}}{\partial r^{\prime}}\right)
-\frac{\partial\mathcal{L}}{\partial r}=0 \label{LagrangEq1}%
\end{equation}
is given by%

\begin{equation}
-4\pi r^{2}\frac{d}{dm}\left[  \rho^{2}\left(  \frac{\partial u}{\partial\rho
}\right)  _{S}\right]  +g=0,\label{LagrangEq2}%
\end{equation}
where $\left(  \partial/\partial\rho\right)  _{S}$ denotes the derivative at a
constant value of the entropy $S$, $g=F/m_{p}$ and the force $F$ affecting the
particle with mass $m_{p}$ is given by [\cite{Verozub2}]%

\begin{equation}
F=-\frac{Gm_{p}M}{r^{2}}\left(1-\frac{r_{g}}{f}\right), \label{Fgrstat}%
\end{equation}
where $f=(r_{g}^{3}$ $+$ $r^{3})^{1/3}.$

Since the pressure is
\begin{equation}
p=\rho^{2}\left(  \frac{du}{d\rho}\right)  _{S}, \label{Pressure}%
\end{equation}
the eq. (\ref{LagrangEq2}) is the equation of hydrostatic equilibrium%

\begin{equation}
\frac{dp}{dr}=\rho g.\label{equilibrium_eq}%
\end{equation}

When the Lagrange equation (\ref{LagrangEq1}) is satisfied, the end condition
can be obtained from the equality%

\begin{equation}
\delta E=\frac{\partial\mathcal{L}}{\partial r^{\prime}}\eta{\Biggl|}_{0}%
^{M}=0, \label{DEltaE}%
\end{equation}
where $\eta=\delta r$. Because of spherical symmetry we have $\eta\left(
0\right)  =0$, and $\eta\left(  R\right)  $ is an arbitrary magnitude, this
equation leads to the following condition at the surface%

\begin{equation}
r^{2}\rho^{2}\left(  \frac{\partial u}{\partial\rho}\right)  _{S}=0.
\label{BounderyCondition1}%
\end{equation}

The second variation $\delta^{2} E$ is of the form%

\begin{equation}
\delta^{2}E=\int\limits_{0}^{M}2\Omega\left(  \eta,\eta^{\prime}\right)
dm,\label{Delta2E}%
\end{equation}
where $2\Omega\left(  \eta,\eta^{\prime}\right)  =\mathcal{L}_{r^{\prime
}r^{\prime}}\eta^{\prime2}+2\mathcal{L}_{rr^{\prime}}\eta\eta^{\prime
}+\mathcal{L}_{rr}\eta^{2}$ , $\eta^{\prime}=d\eta/dm$ and lower indices
denote partial derivatives with respect to $r$ and $r^{\prime}$. The value
of $\delta^{2}E$ is positive only, if the equation%

\begin{equation}
\frac{d}{dm}\left(  \frac{d\Omega}{d\eta^{\prime}}\right)  -\frac{d\Omega
}{d\eta}=\sigma^{2}\eta\label{LagrangEq3}%
\end{equation}
has positive proper values $\sigma^{2}$. Boundary condition is given by%

\begin{equation}
\Omega_{\eta^{\prime}} {\biggl|}_{m = M}=0 . \label{BounderyCondition2}%
\end{equation}

For the functions $\mathcal{L}_{r^{\prime}r^{\prime}}$, $\mathcal{L}%
_{rr^{\prime}}$ and $\mathcal{L}_{rr}$ we find%

\[
\mathcal{L}_{r^{\prime}r^{\prime}}=16\pi^{2}r^{4}\rho^{2}\left[  \rho
^{2}\left(  \frac{\partial^{2}u}{\partial\rho^{2}}\right)  _{S}+2\rho\left(
\frac{\partial u}{\partial\rho}\right)  _{S}\right]  ,
\]%

\begin{equation}
\mathcal{L}_{rr^{\prime}}=8\pi r\rho\left[  \rho^{2}\left(  \frac{\partial
^{2}u}{\partial\rho^{2}}\right)  _{S}+2\rho\left(  \frac{\partial u}%
{\partial\rho}\right)  _{S}-\rho\left(  \frac{\partial u}{\partial\rho
}\right)  _{S}\right]  , \label{Derivatives1}%
\end{equation}%

\[
\mathcal{L}_{rr}=\frac{4}{r^{2}}\left[  \rho^{2}\left(  \frac{\partial^{2}%
u}{\partial\rho^{2}}\right)  _{S}+2\rho\left(  \frac{\partial u}{\partial\rho
}\right)  _{S}-\frac{1}{2}\rho\left(  \frac{\partial u}{\partial\rho}\right)
_{S}\right]  -\frac{\partial g}{\partial r}.
\]

Using (\ref{Pressure}) and equality [\cite{Cox}],[\cite{Shapiro}]%

\begin{equation}
\rho^{2}\left(  \frac{\partial^{2}u}{\partial\rho^{2}}\right)  _{S}%
+2\rho\left(  \frac{\partial u}{\partial\rho}\right)  _{S}=\left(
\frac{\partial p}{\partial\rho}\right)  _{S}=\Gamma\frac{p}{\rho
},\label{ThermoEquation}%
\end{equation}
where $\Gamma$ is the usual generalized adiabatic exponent [\cite{Ledoux}],
that will be identified here with $\Gamma$ in the equation of state%

\begin{equation}
p=K\rho^{\Gamma}\label{adiabata}%
\end{equation}
($K$ is a constant), we obtain
\[
\mathcal{L}_{r^{\prime}r^{\prime}}=16\pi^{2}r^{4}\rho^{2}\Gamma p,
\]%

\begin{equation}
\mathcal{L}_{rr^{\prime}}=8\pi rp\left(  \Gamma-1\right)  ,
\label{Derivative2}%
\end{equation}%

\[
\mathcal{L}_{rr}=\frac{4p}{r^{2}\rho}\left(  \Gamma-\frac{1}{2}\right)
-\frac{\partial g}{\partial r}.
\]

Substituting these magnitudes in (\ref{LagrangEq3}), we obtain following
equation for dimensionless functions $\xi=\eta/r$:%

\begin{equation}
\mathcal{Q}\left(  \xi\right)  =\sigma^{2}\xi, \label{OperatEquation}%
\end{equation}
where
\begin{equation}
\mathcal{Q}(\xi)=-\frac{1}{\rho r^{4}}\frac{d}{dr}\left(  \Gamma p r^{4}%
\frac{d\xi}{dr}\right)  -\frac{1}{\rho r}\left\{  \frac{d}{dr}\left[  \left(
3\Gamma-2\right)  p\right]  +\rho r\frac{\partial g}{\partial r}\right\}  \xi.
\label{OperatorL}%
\end{equation}

The operator $\mathcal{Q}$ is selfconjugate in the Hilbert space.
Its proper values $\sigma_{i}^{2}$ $\left(  i=0,1\ldots\right)  $ form an
infinitely discrete sequence and are orthogonal with the weight $\rho r^{2}$
[\cite{Cox}]. Consequently, if $\xi_{k}$ are the proper function corresponding
to the proper values $\sigma_{k}$, then%

\begin{equation}
\sigma_{k}^{2}=\frac{1}{J_{k}}\int\limits_{0}^{M}\xi_{k}^{\star}\left(
\mathcal{Q}\xi_{k}\right)  r^{2}dm,\label{Sigma2}%
\end{equation}
where $\xi^{\star}$ denotes the complex conjugate magnitude and
\begin{equation}
J_{k}=\int\limits_{0}^{M}\left|  \xi_{k}\right|  ^{2}r^{2}dm.\label{Modul}%
\end{equation}

The numerator of eq. (\ref{Sigma2}) consist of two integral. The first of them
\bigskip$\geq0$ and the second is given by%

\begin{equation}
\int\limits_{0}^{M}\left|  \xi_{k}\right|  ^{2}\left(  -g\right)  \left[
\left(  3\Gamma-4\right)  +\frac{r_{g}\ r^{3}}{\left(  r_{g}^{3}+r^{3}\right)
^{4/3}}\right]  rdm.\label{Integral}%
\end{equation}
Since the magnitude $-g$ and the second term in eq. (\ref{Integral}) are 
positive,
at $\Gamma \geq   4/3$ the integral is positive. For that reason the proper values
$\sigma^{2}>0$. Thus, at least at $\Gamma \geq   4/3,$ the configurations are stable.

\subsection{Numerical Results}

For a numerical solution of eq. (\ref{OperatEquation}) we will rewrite it for
$\Gamma=Const$ in the dimensionless form:
\begin{equation}
\frac{d^{2}\xi}{dx^{2}}=\frac{R}{x\ \lambda_{p}}\left\{  \left(
x-\frac
{4\lambda_{p}}{R}\right)  \frac{d\xi}{dx}+\left[  \frac{\Omega^{2}xGM}%
{R^{2}\Gamma\ g}+\frac{3\Gamma-2}{\Gamma}+\frac{x}{\Gamma\ g}\frac{dg}%
{dx}\right]  \right\}  ,\label{oscillations_eq_1}%
\end{equation}
where $x=r/R$, $\lambda_{p}=-p/\rho g$, $\Omega=\sigma
(R^{3}/GM)^{1/2}$ is dimensionless angular frequency, and the
function $\rho(x)$ is the solution of the equation of hydrostatic equation
(\ref{equilibrium_eq}).

Since $\lambda_{p}/R\rightarrow0$ at $x\rightarrow1$, the magnitude in the
figure brackets must tend to zero when $x\rightarrow1$. Therefore, on the
surface of the object the following equation is valid%

\begin{equation}
\frac{d\xi}{dx}\left( 1 \right)=\left( -
\frac{\Omega^{2}GM}{R^{2}\Gamma\ g}-\frac
{3\Gamma-2}{\Gamma}-\frac{dg/dx}{\Gamma\ g}\right) _{x=1}.
\end{equation}
Besides that, in the case of spherical symmetry, $\xi(0)=0$ and we set
$\xi(1)=1$.

\begin{figure}
\resizebox{115mm}{!}
{\includegraphics[width=115mm,height=60mm]{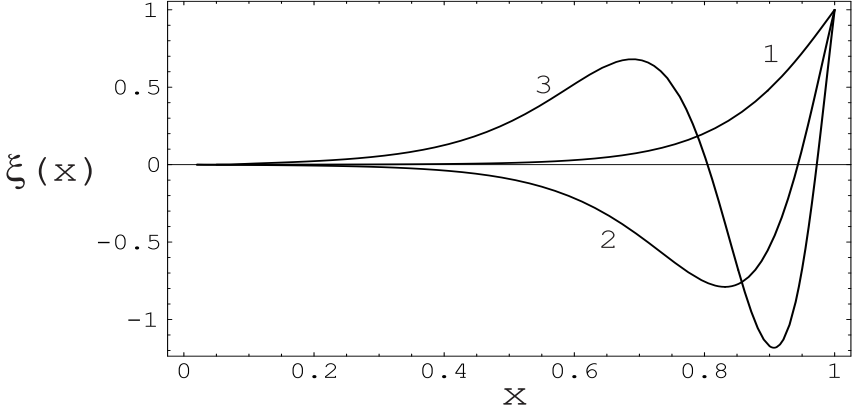}}
\hfill
\parbox[b]{55mm}{\caption{The
functions $\xi(x)$ for the first three modes of the oscillations of the
configuration with mass $2.6\cdot10^{6}M_{\odot}$}
\label{oscillation_eq_1}}
\end{figure}

Consider, for example, the configuration with mass
$M=2.6\cdot10^{6}M_{\odot}$. Setting in eq. (\ref{adiabata})
$\Gamma=5/3$ and $K=10^{13}$ (in CGS units), we find by numerical
methods for  eq. (\ref{oscillations_eq_1}) the three first proper
functions $\xi(x)$ that are shown in Fig. \ref{oscillation_eq_1}.

The periods of these oscillations are given by:
\begin{equation}
T_{1}=51s,\ T_{1}=14s,\ T_{3}=6.7s.
\end{equation}

\newpage
%
%

\addresses
\rf{L.Verozub, Dept. of Physics and Astronomy, Kharkov National University,
Kharkov, 61077, Ukraine,\\ e-mail: verozub@gravit.kharkov.ua} \rf
{A.Kochetov, Dept. of Physics and Astronomy, Kharkov National University,
Kharkov, 61077, Ukraine,\\ e-mail: alex@gravit.kharkov.ua}
\end{document}